\documentclass[12pt]{iopart}
\usepackage{iopams}
\usepackage{graphicx}
\usepackage{bm}
\usepackage{gensymb}
\usepackage{color}
\begin{document}
\title{Decoherence in an infinite range Heisenberg model}
\author{A. Dey$^*$, M. Q. Lone$^*$ and S. Yarlagadda}
\address{TCMP Div.,
1/AF Salt Lake, Saha Institute of Nuclear physics, Kolkata, India.}
\ead{y.sudhakar@saha.ac.in}
\date{\today}

\begin{abstract} 
We study decoherence  in an  infinite range Heisenberg model (IRHM) 
in the two situations where the system is coupled to a bath of either local optical phonons or global optical phonons. 
 Using a non-perturbative treatment, we derive an effective Hamiltonian that is valid in the regime of
 strong  spin-phonon coupling under non-adiabatic conditions. 
It is shown that the effective Hamiltonian commutes with the IRHM
 and thus has the same eigenstates as the IRHM.
 By analyzing the dynamics of the system using a quantum master equation approach, 
we show that the quantum states of the IRHM system do not decohere under  Markovian dynamics
when the spins interact with local phonons. 
{For interactions with global phonons, 
 the off-diagonal matrix elements of the
system's reduced density matrix,
obtained for non-Markovian dynamics,
 do not indicate decoherence only when states with the
same $S^z_T$ (i.e., eigenvalue for the z-component
of the total spin) are considered.}
\end{abstract}
\maketitle
\section{Introduction} 
Quantum information processing heavily relies on  a precious and fragile 
resource, namely,  
quantum entanglement \cite{6}.
The  fragility
of entanglement is due to the coupling between a quantum system and its environment;
such a coupling leads to decoherence, the process
 by which information is degraded.
Decoherence is  the fundamental mechanism by which fragile superpositions are destroyed
thereby producing 
a quantum to 
classical transition \cite{schloss,zurek2}. 
In fact, decoherence is one of the main obstacles for the preparation, observation, and
implementation of multi-qubit entangled states.
The intensive work on quantum information and computing in recent years has 
tremendously increased the
interest in exploring and controlling decoherence effects   \cite{nat1,milb2,QA,CJ,zurek,diehl,verst,weimer}.

Since coupling of a quantum system
to the environment and the concomitant entanglement fragility are 
ubiquitous
 \cite{6,schloss}, it is imperative that progress be made in
minimizing decoherence.
Decoherence free states
prevent the loss of information due to destructive environmental interactions
and thus circumvent the need for stabilization methods for
quantum computation and quantum information. 
{ In the past decoherence-free-subspace (DFS)
\cite{palma, zanardi}
has been shown to exist in the Hilbert space of
a model 
where all qubits of the quantum system 
are coupled to a common environment
with equal strength. 
A DFS is a
subspace which is invariant under the action of the system
Hamiltonian; furthermore, the subspace is spanned by degenerate eigenvectors of the
system operators coupling to the environment 
\cite{whaley2, whaley3}.
Alternately, decoherence  can also be suppressed through quantum control strategies \cite{hohen,lidar,goan2}.}

{
Although the theory of decoherence has undergone major advances \cite{schloss,zurek2}, 
yet, there exist many definitions of decoherence \cite{coles}.
For the analysis in this paper, we choose the most commonly used
definition of decoherence:  Loss of off-diagonal elements
in the system's reduced density matrix.
In general, a many-qubit (i.e., many-spin) system can have 
distance dependent interaction. The two limiting cases for interaction are
spin interactions that are independent of distance and spin chains with nearest-neighbor
interactions only.
In this work we consider
the extreme case of distance independent interaction
among the spins, i.e., the IRHM.} 
The objective of this paper is to study the decoherence 
phenomenon, due to coupling of spins of the IRHM to optical phonons, 
in the two extreme cases of the spins
being independently coupled to different baths; and
all the spins being collectively coupled to the same environment.
{
{ We employ 
the analytically simpler frame of reference
of hard-core-bosons (HCBs) rather than that of spins so that the single particle excitation spectrum can be easily 
obtained and exploited;} we show that the effective
Hamiltonian 
even in higher order (i.e., greater than second order)  perturbation theory
retains the same eigenstates as
the IRHM when the spins are coupled to local phonons. Furthermore, decoherence is studied using
 the quantum master equation approach {\cite{nazir}}. Our dynamical analysis 
 shows that the system  coupled to local phonons does not decohere  
when Markov processes are considered; whereas for global phonons,
 even for non-Markovian dynamics,
there is no decoherence when eigenstates with
the same eigenvalue $S^z_T$ (i.e., z-component of the total spin) 
are considered.}

The rest of the paper is organized as follows:
In section $2$, we introduce the IRHM Hamiltonian and describe its eigenstates and eigenenergies. 
In section $3$,
we  study decoherence under strong coupling with 
local optical phonons,
and show that the effective Hamiltonian thus obtained 
 retains the same eigenstates as $H_{\rm IRHM}$. In section $4$, 
we use the master equation approach 
and show that the system does not decohere under 
 local and global
 couplings.
 Next, in section $5$, we give our conclusions and make some general remarks
regarding the wider context of our results. 
The paper also contains an appendix where we derive the third order perturbation
contribution to
our effective Hamiltonian ($H_{eff}$) and show that the eigenstates of the IRHM Hamiltonian are retained
by our $H_{eff}$.

\section{Infinite Range Heisenberg Model}

We begin by introducing the IRHM whose decoherence will be studied
when the system is coupled to
 either local or global optical phonons.
The IRHM is defined as:
\begin{eqnarray}
\!\!\!\!\!\!\!\! H_{\rm IRHM} \!&=& \!J\! \sum_{i,j>i} \!\! \left [ \vec{S_i}.\vec{S_j} + (\Delta -1)S^z_i S^z_j \right] 
\nonumber \\
\!&=& \!\frac{J}{2} \! \left [ \!
 \left ( \sum_{i} \vec{S_i} \right )^2 \!\!
 - \! \left ( \sum_{i} \vec{S_i}^2  \right ) \right . 
\nonumber \\
\!&& ~ + \left . (\Delta -1)\left \{ \left ( \sum_i S^z_i \right )^2 
- \left ( \sum_i {S^{z}_i}^2\right ) 
\right \}
 \!\right ] ,
\label{H_gen}
\end{eqnarray} 
where $J > 0$, $\Delta \geq 0$, and we are considering only $S=1/2$ spins.
We note that $H_{\rm IRHM}$
 commutes with both  $S^z_{Total}$ ($\equiv \sum_i S^z_i$) and $\left ( \sum_{i} \vec{S_i} \right )^2$ ($\equiv S^2_{Total}$).
In equation (\ref{H_gen}), it is understood that $J = J^{\star}/(N-1)$ 
(where $ J^{\star}$ is a finite
quantity) so that the energy per site remains finite as $N \rightarrow \infty$.
The eigenstates of $H_{\rm IRHM}$ are characterized by $S_T$ (i.e., the total spin eigenvalue)
and $S^z_T$ (or the eigenvalue of the z-component of the total spin $S^z_{Total}$); the 
  eigenenergies of these eigenstates are
\begin{eqnarray}
\!\!\!\! E_{S_T} = \frac{J}{2} \left [ S_T (S_T + 1) - \frac{3N}{4} + (\Delta -1) \left ({S^z_T}^2 -\frac{N}{4} \right )\right ] .
\end{eqnarray}
 The ground state corresponds to $S^z_T =0$ and $S_T=0$ which is rotationally
invariant.

\section{Effective Hamiltonian for IRHM spins coupled to local optical phonons}

The real quantum computer will not be free from noise
and thus the entangled states have a tendency to undergo decoherence.
To study decoherence due to phonons, we consider interaction with
 optical phonons such as would
be encountered when considering transition metal oxides.
We will now derive an effective Hamiltonian, when the spins of the IRHM are coupled to local
optical phonons, and show that
the eigenstates of the effective Hamiltonian are the same as those of $H_{\rm IRHM}$.

 The total Hamiltonian $H_T$ is given by
\begin{eqnarray}
\!\!\!\! H_T = H_{\rm IRHM}+ g \omega\sum_i S^z_i (a^{\dagger}_i + a_i) + \omega \sum_i a^{\dagger}_i a_i ,
\label{Ham3}
\end{eqnarray}
where $a$ is the phonon destruction operator \cite{indexk}, $\omega$ is the optical phonon frequency, and $g$ is the
coupling strength. Now, we make the connection that the spin operators can be expressed in terms of
HCB creation and destruction operators $b^{\dagger}$ and $b$, i.e., 
$b^{\dagger} = S^{+}$, $b = S^{-}$, and $b^{\dagger} b = S^{z} + 0.5$. We then
observe that conservation of $S^z_{Total}$ implies conservation of total number of HCB. 
The total Hamiltonian is then given by
\begin{eqnarray}
H &=&  J \sum_{i,j>i}[(0.5 b^{\dagger}_i  b_{j} +{\rm H.c.}) + \Delta (n_i - 0.5)(n_j -0.5)  ] 
\nonumber \\
          &&+ \omega \sum_j a^{\dagger}_{j} a_j
          + g \omega \sum_j (n_j-\frac{1}{2}) 
 (a_j +a^{\dagger}_j) , 
\end{eqnarray}
where
 $n_j \equiv b^{\dagger}_{j} b_j $.
Subsequently, we perform 
the well-known Lang-Firsov (LF) transformation \cite {lang,sdadys} on this
Hamiltonian. 
Under the LF transformation  given by $e^S H e^{-S} =H_0+H_I $ with
$S= - g \sum_i (n_i-\frac{1}{2}) (a_i - a^{\dagger}_i)$, the operators
$b_j$ and $a_j$ transform like fermions  and bosons;
this is due to the interesting 
 commutation properties of HCB given below: 
\begin{eqnarray}
[b_i,b_j]&=&[b_i,b^{\dagger}_j]= 0 , \textrm{ for } i \neq j , \nonumber\\
\{b_i,b^{\dagger}_i\}& = & 1 .
\label{commute}
\end{eqnarray}   
Next, the unperturbed Hamiltonian $H_0$ is expressed as 
\cite{sdadys}
\begin{eqnarray}
\!\!
H_0 =  H_s +H_{env} ,
\label{H0}
\end{eqnarray}
where  we identify $H_s$ as the system Hamiltonian
\begin{eqnarray}
\!\!\!\! H_s &=& J \sum_{i,j>i}[(0.5 e^{-g^2}b^{\dagger}_i  b_{j} +{\rm H.c.})
\nonumber \\
&& ~~~~~~~~~~ + \Delta (n_i -0.5)( n_j - 0.5)  ] ,
\label{Hs}
\end{eqnarray}
and $H_{env}$ as the Hamiltonian of the environment
\begin{eqnarray}
H_{env} = \omega \sum_j a^{\dagger}_j a_j . 
\end{eqnarray}
On the other hand, the interaction $H_I$ which we will treat as perturbation is given by
\begin{eqnarray}
H_I
= J \sum_{i,j>i}[0.5 e^{-g^2}b^{\dagger}_i  b_{j} ]
            \{\mathcal S^{{ij}^\dagger}_+ \mathcal S^{ij}_{-}-1\} +{\rm H.c.} ,
\label{int}
\end{eqnarray}
where $\mathcal S^{ij}_{\pm} = \textrm{exp}[\pm g(a_i - a_{j})]$.
 { In the transformed frame, the system Hamiltonian depicts that all the HCBs are coupled to the same
phononic mean-field. Thus, the unperturbed Hamiltonian $H_0$ comprises of
the system Hamiltonian $H_s$ representing
  HCBs with the same reduced hopping term $0.5 J e^{-g^2}$
and the environment Hamiltonian $H_{env}$ involving displaced bath oscillators corresponding to local distortions.
Here it should be pointed out that both the 
interaction of the HCB with the mean-field as well as the local polaronic distortions in the bath oscillators
  involve controlled degrees of freedom. { Now, the system Hamiltonian $H_s$ can be expressed as} 
  \begin{equation}
   H_s=H_{\rm IRHM}+(H_s-H_{\rm IRHM})
  \end{equation}
  {When we change the Hamiltonian from  $H_{\rm IRHM}$ to $H_s$ by adiabatically turning on the perturbation $(H_s-H_{\rm IRHM})$,}
the resulting state of the system is still obtainable from that of $H_{\rm IRHM}$ by using unitary
Hamiltonian dynamics 
  and is thus predictable based on a knowledge of the coupling parameter $g$ \cite{gl}.
Thus  no irreversibility
is involved in going from $H_{\rm IRHM}$ to $H_s$.
 {\it On the other hand, perturbation $H_{I}$ pertains to the interaction of HCBs with local 
deviations from the phononic mean-field;
  the interaction term $H_I$ represents numerous
or uncontrolled environmental degrees of freedom and thus has the potential
for producing decoherence}. Furthermore, it is of interest to note that
the interaction term is weak in the transformed frame
compared to the interaction 
in the original frame; thus one can perform perturbation theory with the interaction term.
}

 We represent the  eigenstates of the unperturbed Hamiltonian $H_{0}$ as
$|n,m\rangle\equiv|n\rangle_{s}\otimes|m\rangle_{ph}$ with the corresponding
 eigenenergies $E_{n,m}=E_{n}^{s}+E_{m}^{ph}$;
$|n\rangle_{s}$ is the eigenstate of the system with eigenenergy $E_{n}^{s}$
while $|m\rangle_{ph}$ is the eigenstate for  the environment with eigenenergy $E_{m}^{ph}$.
 Henceforth, for brevity, we will use $\omega_m \equiv E_m^{ph}$.
On observing that $\langle0,0|H_{I}|0,0\rangle=0$ ({ i.e., the ground state
expectation value of the deviations is zero}),
we obtain the next relevant second-order perturbation term \cite{sdadys}
\begin{eqnarray}
E^{(2)}=\sum_{n,m}{{\langle0,0|H_{I}|n,m\rangle\langle n,m|H_{I}|0,0\rangle}\over{E_{0,0}-E_{n,m}}} .
\end{eqnarray}
{ For strong coupling ($g>1$) and non-adiabatic ($J^{\star}/\omega \leq 1$) conditions,
on noting that 
$\omega_{m}-\omega_{0} =\omega_m$ is a positive integral multiple of $\omega$  and that
$E_{n}^{s}-E_{0}^{s}
\sim J^{\star}e^{-g^2} \ll \omega$ (as shown in the next section), we 
get the following second-order term $H^{(2)}$ \cite{sdys}
using Schrieffer-Wolff (SW) transformation (as elaborated in
Appendix A of references \cite{srsypbl} and \cite{srsypbl2}):}
\begin{eqnarray}
\!\!\!\!H^{(2)} \! &=& \!
-\sum_{m}{{{_{ph}\langle0|H_{I}|m\rangle_{ph}}~{_{ph}\langle m|H_{I}|0\rangle_{ph}}}\over{\omega_{m}}} 
\nonumber \\
&=& \! \sum_{i, j > i } 
 \left [(0.5 J_{\perp}^{(2)}
b^{\dagger}_i
 b_j +{\rm H.c.}) \right .
\nonumber \\
&&~~~~~ \left . - 
 0.5 J_{\parallel}^{(2)}
 \{n_i(1-n_j)+n_j(1-n_i)\} \right ] ,
\label{H_eff}
\end{eqnarray}
where 
\begin{eqnarray}
\!\!\!\!\!\!\!\!\!\!\!\! J_{\perp}^{(2)} \equiv  - (N-2) f_1 (g) \frac{J^2 e^{-2g^2}}{2 \omega}
 \sim -(N-2) \frac{J^2e^{-g^2}}{2g^2 \omega} ,
 \end{eqnarray}
\begin{eqnarray}
 J_{\parallel}^{(2)} \equiv 
[2f_1 (g)+f_2(g)]\frac{J^2 e^{-2g^2}}{{2 \omega}} 
\sim \frac{J^2 }{{4g^2 \omega}}, 
 \end{eqnarray}
 with
$f_1(g) \equiv \sum^{\infty}_{n=1} g^{2n}/(n!n)$
 and
$f_2(g) \equiv \sum^{\infty}_{n=1}\sum^{\infty}_{m=1} g^{2(n+m)}/[n!m!(n+m)]$. 
{The effective Hamiltonian $H_{s}+H^{(2)}$ is a low energy Hamiltonian
obtained by the canonical SW transformation \cite{schrieffer,loss2} decoupling the low-energy and the
high-energy subspaces; this decoupling  is a consequence of $J^{\star}e^{-g^2} \ll \omega$. }
We now make the important observation that 
the effective Hamiltonian $H_{s}+H^{(2)}$, when expressed in terms of spins, 
 has the following form:
\begin{eqnarray}
 \sum_{i, j > i } 
 \left [ J_{\rm tr} ({S_i^{x}}{S_j^{x}} + {S_i^{y}}{S_j^{y}}) + J_{\rm lng} S_i^zS_j^{z} \right ] ,
\label{Heff}
\end{eqnarray}
and thus has eigenstates identical to
those of the original Hamiltonian $H_{\rm IRHM}$ in equation (\ref{H_gen})
because $\sum_{i, j > i } 
 (S_i^{z} S_j^{z})$ and $ H_{\rm IRHM}$ commute. 
On carrying out higher order 
(i.e., beyond second order) perturbation theory (as discussed in Appendix A),
and expressing the results in the spin language,
we still get an effective Hamiltonian $H_{eff}$ of the following form that
has the same eigenstates as the IRHM.
\begin{eqnarray}
H_{eff} \! &=& \! \sum_{i, j > i } 
 \left [ J_{xy}  ( \sum_k S_k^z ) ({S_i^{x}}{S_j^{x}} + {S_i^{y}}{S_j^{y}}) \right ]
\nonumber \\
 &&+
 \sum_i J_{z}  (\sum_k S_k^z  )S_i^{z} ,
\label{Heff}
\end{eqnarray} 
where $J_{xy}$ and $J_z$ are functions of the $S^z_{Total}$  ($= \sum_k S_k^z$ ) operator.
The small parameter of our perturbation theory, for a small N system, is $J/(g \omega)$ [see reference \cite{sy1} for details];
whereas for a large N, the small parameter is $J^{\star}/(g^2 \omega)$ [see reference \cite{sm_par} for an explanation].
It is the infinite range of the Heisenberg model that enables  
 the eigenstates of the system to 
remain unchanged.
Next, we study decoherence in a dynamical
context and gain more insight into how the states
of our $H_{\rm IRHM}$ can be decoherence free.

{\section{Dynamical evolution of the system}}
In this section,
we will study decoherence in the system from the dynamical perspective.
We will discuss the dynamics of an
open quantum system, described by the Hamiltonian IRHM, using master equation approach. 
Our quantum system is open because it is coupled to
another quantum system, i.e., a bath or environment \cite{Pet}. 
In our case, IRHM is coupled to a bath of either local optical phonons
{ [see equation (\ref{Ham3})]
or global optical phonons [see equation (\ref{Ham2})].}
As a consequence of the system-environment  coupling, 
the state of the system  may change.  This interaction may lead to 
certain system-environment correlations such that
the resulting state of the system
may  no longer be represented in terms of unitary Hamiltonian dynamics. 
The dynamics of the system, described by the reduced density matrix $\rho_s(t)$ at 
time $t$, is obtained from  the density matrix $\rho_T(t)$ of the total system by taking the 
partial trace over the degrees of freedom of the environment:
\begin{eqnarray}
\rho_s(t) = Tr_R\left[\rho_T(t)  \right]= Tr_R\left[  U(t) \rho_T(0) U^{\dagger}(t)     \right] ,
\end{eqnarray}
where $U(t)$  represents the time-evolution operator of the total system. Now it is
evident from the above equation that we need first to determine 
the dynamics of the total system which is a  difficult task 
in most of the cases. 
By contrast, master equation approach 
conveniently  and directly yields the time evolution of
the reduced density matrix 
 of the system interacting  with an environment.
This approach relieves us from the need of having to first determine the dynamics of the
total system-environment combination and then to trace out the degrees
of freedom of the environment.\\

\subsection{Decoherence due to Local Optical Phonons:}
We begin this sub-section by considering the following Hamiltonian:
\begin{eqnarray}
H=  H_0 + H_I ,
\end{eqnarray}
where $H_0$ is the system-environment Hamiltonian given by equation (\ref{H0}) 
and $H_I$ represents the interaction Hamiltonian given by equation (\ref{int}).
It is convenient and simple to derive the quantum master equation
in the  interaction picture. Thus our starting point is the interaction picture
von Neumann equation for the total density operator $\tilde{\rho}_T(t)$ 
\begin{eqnarray}
\frac{d \tilde{\rho}_T (t)}{dt} = -i [\tilde{H_I}(t), \tilde{\rho}_T(t)] ,
\label{von1}
\end{eqnarray} 
where $\tilde{H}_I(t) = e^{iH_ot} H_I e^{-iH_ot} $ and 
$\tilde{\rho}_T (t) = e^{iH_ot} \rho_T(t) e^{-iH_ot} $ are the interaction
Hamiltonian and the total system density matrix operators (respectively) 
expressed in the interaction picture.
 Re-expressing the above equation in integral form yields
\begin{eqnarray}
\tilde{\rho}_T(t) = \tilde{\rho}_T(0) - i \int_{0}^t d \tau [\tilde{H_I}(\tau), \tilde{\rho}_T(\tau)].
\label{von2}
\end{eqnarray}

{ Nowadays there is considerable interest in systems with initial correlation with the environment \cite{modi,morozov};
however, for simplicity, 
let us suppose that the initial state of the 
total system is a factorized state given as
$ \rho_T(0) = \rho_s(0) \otimes R_0$  with  
{ $R_0=\sum_{n}|n \rangle_{ph}~\!\! _{ph}\langle n|e^{-\beta \omega_n}/Z$}
being the initial thermal density matrix operator of the environment
 and $\beta = \frac{1}{k_B T}$;} 
furthermore, 
$Z=\sum_n e^{-\beta \omega_n}$
 defines the partition function of the environment. 
With this assumption, we 
substitute  equation (\ref{von2}) inside the commutator of
 equation (\ref{von1}) and  then take the trace over the environmental degrees 
of freedom to obtain the following equation:
\begin{eqnarray}
\!\!\!\!\!\!\!\!\!\!
\frac{d \tilde{\rho}_s(t)}{dt} &=& 
-i~Tr_R[\tilde{H}_I(t), \tilde{\rho}_s(0) \otimes R_o] \nonumber \\
&-& \int_0^t d\tau Tr_R[\tilde{H}_I(t),[\tilde{H}_I(\tau), \tilde{\rho}_T(\tau)]] .
\label{mas1}
\end{eqnarray}
The 
above equation  still contains 
the total density matrix $\tilde{\rho}_T(\tau)$; In order to  evaluate
it, we rely on an approximation known as the Born approximation. This approximation
assumes that the environment degrees of freedom are large and thus the effect on
the environment due to the system is negligibly small for a weak system-environment coupling.
As a consequence, we write   $\tilde{\rho}_T(\tau) = \tilde{\rho}_s(\tau) 
\otimes R_0 + \mathcal{O}(\tilde{H_I})$ 
within the second order perturbation in system-environment interaction \cite{Pet,HJ,CM,YJ1,YJ2,HFPB,HPB,MS,EF,goan}.
 Therefore we can write the equation (\ref{mas1}) in time-local form as
\begin{eqnarray}
\!\!\!\!\!\!\!\!\!\!
\frac{d \tilde{\rho}_s(t)}{dt} &=& 
-i~Tr_R[\tilde{H}_I(t), \rho_s(0) \otimes R_o] \nonumber \\
&-& \int_0^t d\tau Tr_R[\tilde{H}_I(t),[\tilde{H}_I(\tau), \tilde{\rho}_s(t) \otimes R_0]] .
\label{mas2}
\end{eqnarray}
We note here that, for obtaining the non-Markovian time-convolutionless  master equation (\ref{mas2}),
we replaced $\tilde{\rho}_s(\tau) $ with $\tilde{\rho}_s(t)$. This replacement
is equivalent to obtaining a time-convolutionless master equation perturbatively 
up to only second order in the interaction Hamiltonian using the 
time-convolutionless projection operator technique \cite{Pet, HFPB, HPB}.
It has been shown in a number of cases 
that time-local approach works better than
time-nonlocal approach \cite{Pet, YJ1, MS, EF, UK}.
{ Now we will consider the 
second order time-convolutionless master equation (\ref{mas2}) with
the time variable $\tau$ replaced 
by $(t - \tau)$.
\begin{eqnarray}
\!\!\!\!\!\!\!\!\!\!
\frac{d \tilde{\rho}_s(t)}{dt} &=& 
-i~Tr_R[\tilde{H}_I(t), \rho_s(0) \otimes R_o] \nonumber \\
&-& \int_0^t d\tau Tr_R[\tilde{H}_I(t),[\tilde{H}_I(t-\tau), \tilde{\rho}_s(t) \otimes R_0]] .
\label{mas3}
\end{eqnarray}
 Next, we will study the Markovian dynamics of the system. To this end we assume that the correlation 
time scale  $\tau_{c}$  for the environmental fluctuations 
is negligibly small compared to the relaxation time scale $\tau_{s}$  for the system,
i.e.,  $\tau_{c} \ll \tau_{s}$. This time scale assumption is 
 {motivated}
by the condition 
 $J^{\star}e^{-g^2} \ll \omega$ already mentioned in section $3$. 
The Markov approximation ($\tau_{c} \ll \tau_{s}$)
allows us to set the upper limit of the integral to $\infty$ in equation (\ref{mas3}).
 Thus we obtain the second order time-convolutionless
 Markovian master equation (\ref{mark}):}
\begin{eqnarray}
\!\!\!\!\!\!\!\!\!\!
\frac{d \tilde{\rho}_s(t)}{dt} &=& 
-i~Tr_R[\tilde{H}_I(t), \rho_s(0) \otimes R_o] \nonumber \\
&-& \int_0^\infty d\tau Tr_R[\tilde{H}_I(t),[\tilde{H}_I(t - \tau), \tilde{\rho}_s(t) \otimes R_0]] .
\label{mark}
\end{eqnarray}
Defining $\{|n\rangle_{ph}\}$ as the basis set for phonons, therefore, we can write
the master equation as:
{\begin{eqnarray}
\!\!\!\!\!\!\!\!\!\!\!
\frac{d \tilde{\rho}_s(t)}{dt} &=&-i  \sum_{n}~ _{ph}\langle n| [\tilde{H}_I(t), \rho_s(0) \otimes R_o] |n \rangle_{ph}
\nonumber \\
&& - \sum_n \int_0^\infty d\tau \left[~ _{ph}\langle n |
 \tilde{H}_I(t)\tilde{H}_I(t-\tau)\tilde{\rho}_s(t) \otimes R_o|n\rangle_{ph}  \right .
 \nonumber \\
&& ~~~~~~~~~~~~~~~~ - ~_{ph}\langle n| \tilde{H}_I(t)\tilde{\rho}_s(t) \otimes R_o \tilde{H}_I(t-\tau) |n\rangle_{ph} \nonumber \\
&& ~~~~~~~~~~~~~~~~ - ~_{ph}\langle n| \tilde{H}_I(t-\tau)\tilde{\rho}_s(t) \otimes R_o \tilde{H}_I(t) |n\rangle_{ph}
 \nonumber \\
&& ~~~~~~~~~~~~~~~~
 + \left .
~_{ph}\langle n | \tilde{\rho}_s(t) \otimes R_o \tilde{H}_I(t-\tau)\tilde{H}_I(t)|n\rangle_{ph}
 \right] .
\label{mas}
\end{eqnarray}}

In order to simplify the above master equation, we need to evaluate  
the time evolution of the operators involved in $H_I$. Considering the 
second term in the equation (\ref{mas}), yields
{\begin{eqnarray}
\fl _{ph}\langle n |
 \tilde{H}_I(t)\tilde{H}_I(t-\tau)\tilde{\rho}_s(t) \otimes R_o|n\rangle_{ph} 
\nonumber \\
\fl ~ =\sum_m e^{iH_s t} {_{ph}\langle n | H_I |m\rangle_{ph}} e^{-iH_s t} ~
    e^{iH_s (t-\tau)}  {_{ph}\langle m| H_I|n\rangle_{ph}} e^{-iH_s( t-\tau)} \tilde{\rho}_s(t)
\frac{e^{-\beta \omega_n}}{Z} e^{i(\omega_n-\omega_m)\tau} .
\label{time}
\end{eqnarray}}

In momentum space, we express HCB operators as:
$ b^{\dagger}_j =
 \frac{1}{\sqrt{N}}\sum_k e^{ikr_j }~ b^{\dagger}_k$ and $ b_j = \frac{1}{\sqrt{N}}\sum_k e^{-ikr_j }~ b_k$; 
then, it is important to note that the hopping term 
in the system Hamiltonian can be written as:
\begin{eqnarray}
0.5 J \sum_{i,j>i}( e^{-g^2} b^{\dagger}_i  b_{j} +{\rm H.c.}) &=& 
0.5 J e^{-g^2} \left[ \sum_{i, j}b^{\dagger}_i  b_{j} - \sum_{i}b^{\dagger}_i  b_{i} \right] \nonumber \\
&=& 0.5 J^{\star}(\frac{N}{N-1}) e^{-g^2} \hat{n}_0 - 0.5 J e^{-g^2} \hat{N_p}  \nonumber \\
&=&\sum_{k} \epsilon_k b^{\dagger}_k b_k , 
\end{eqnarray}
where we used $ J = J^{\star}/(N-1)$,  $\hat{N_p}\equiv \sum_k b^{\dagger}_k b_k$ and 
 $ \hat{n}_0 \equiv b^{\dagger}_0 b_0$ (i.e., 
the particle number in momentum $k=0$ state). 
{Here it should be mentioned that using HCBs instead of spins has enabled us to
obtain (with ease) the excitation spectrum $\epsilon_k$ which is crucial
for the analysis given below}.
Let $\{ |q \rangle_s\}$  denote the  complete set of 
energy eigenstates (with eigenenergies $E_q^s$) of the system Hamiltonian $H_s$;
 then we can write:
{\begin{eqnarray}
\fl e^{iH_s t} H_I  e^{-iH_s t} 
= 0.5 J e^{-g^2}\sum_{l,j>l}\sum_{q, q^{\prime}} |q \rangle_s {_s \!\langle} q| 
 e^{iH_st} \left[  \frac{1}{N}  
\sum_{k, p}b^{\dagger}_k b_p e^{i(k r_l - p r_j)} \right] e^{-iH_st} |q^{\prime} \rangle_s {_s\!\langle} q^{\prime} |  
  \{\mathcal S^{{lj}^\dagger}_+ \mathcal S^{lj}_{-}-1\}  \nonumber \\
 ~~~~~~~~~~~~~~~~~~~~~~~~~~~~~~~~~~~~~~~~~~~~~~~~~~~~~~~~+ {\rm H. c.} ,
  \nonumber \\
\end{eqnarray}}
which implies
\begin{eqnarray}
 e^{iH_s t}~ _{ph}\langle n | H_I |m\rangle_{ph} e^{-iH_s t}
 = 
 \sum_{q, q^{\prime}} |q \rangle_s {_s\!\langle} q| ~ _{ph}\langle n | H_I |m\rangle_{ph} 
 |q^{\prime} \rangle_s {_s\!\langle} q^{\prime} | e^{i(E_q^s-E_{q^{\prime}}^s)t} ,
 \nonumber \\
\label{TE}
\end{eqnarray}
where $|E_q^s-E_{q^{\prime}}^s| = 0.5 J^{\star} (\frac{N}{N-1}) e^{-g^2}$ or $0$ . {Here we have taken  
 the total number of HCBs to be conserved; then,
 only the hopping term in $H_s$ will contribute to the particle excitation energy.}
Substituting equation (\ref{TE}) in equation (\ref{time}), we get
{\begin{eqnarray}
\fl _{ph}\langle n |
 \tilde{H}_I(t)\tilde{H}_I(t-\tau)\tilde{\rho}_s(t) \otimes R_o|n\rangle_{ph} 
\nonumber \\
\fl =
 \sum_m \sum_{q, q^{\prime}, q^{\prime \prime}} \left [ |q \rangle_s {_s\!\langle} q| ~ _{ph}\langle n | H_I |m\rangle_{ph} 
|q^{\prime} \rangle_s {_s\!\langle} q^{\prime} |~ _{ph}\langle m | H_I |n\rangle_{ph}
 |q^{\prime \prime} \rangle_s {_s\!\langle} q^{\prime\prime} | 
e^{i[(E_q^s-E_{q^{\prime}}^s)t+(E_{q^{\prime}}^s-E_{q^{\prime\prime}}^s)(t-\tau)]}   \right]
\nonumber \\
~~~~~~~~~~~~~~~~~~~~~~~~~~~~~~~~~~ \times  \tilde{\rho}_s(t)\frac{e^{-\beta \omega_n}}{Z} e^{i(\omega_n-\omega_m)\tau} .
\nonumber \\ 
\label{time1}
\end{eqnarray}}
 Thus under the assumption of $J^{\star} e^{-g^2} < < \omega$,
it follows that $|\omega_n - \omega_m| > > |E_q^s - E_{q^{\prime}}^s|$ and
$|\omega_n - \omega_m| > > |E_{q^{\prime}}^s - E_{q^{\prime \prime}}^s|$;
hence in equation (\ref{time1}),  we can 
take 
$e^{i[(E_q^s-E_{q^{\prime}}^s)t]} =1$
 and
$e^{i[(E_{q^{\prime}}^s-E_{q^{\prime\prime}}^s)(t-\tau)]} =1$
 which implies that 
{
we do not get terms
producing decay}. The resultant equation is
\begin{eqnarray}
\!\!\!\!\!\!\!
\fl _{ph}\langle n |
 \tilde{H}_I(t)\tilde{H}_I(t-\tau)\tilde{\rho}_s(t) \otimes R_o|n\rangle_{ph} =
\sum_m {_{ph}\langle n | H_I |m\rangle_{ph}} ~
     {_{ph}\langle m| H_I|n\rangle_{ph}} ~ \tilde{\rho}_s(t) 
 \frac{e^{-\beta \omega_n}}{Z} e^{i(\omega_n-\omega_m)\tau} .
 \nonumber \\
\label{time2}
\end{eqnarray}

Carrying out the same analysis on the remaining (i.e., third, fourth, and fifth) terms in the 
master equation, we write equation (\ref{mas})
as:
{\begin{eqnarray}
\fl ~~~~~\frac{d \tilde{\rho}_s(t)}{dt} = -i  \sum_{n}~ _{ph}\langle n| [\tilde{H}_I(t), 
\tilde{\rho}_s(0) \otimes R_o] |n \rangle_{ph}
\nonumber \\
 - \sum_{n,m} \int_0^\infty d\tau \left[ _{ph}\langle n | H_I |m\rangle_{ph}~
_{ph}\langle m| H_I|n\rangle_{ph} ~ \tilde{\rho}_s(t) 
 \frac{e^{-\beta \omega_n}}{Z} e^{i(\omega_n-\omega_m)\tau} \right.
 \nonumber \\
 ~~~~~~~~~~~~~~~~ - ~_{ph}\langle n | H_I |m\rangle_{ph} ~ \tilde{\rho}_s(t) ~ _{ph}\langle m|
 H_I|n\rangle_{ph} \frac{e^{-\beta \omega_m}}{Z} e^{i(\omega_n-\omega_m)\tau}
\nonumber \\ 
~~~~~~~~~~~~~~~~ - ~ _{ph}\langle n | H_I |m\rangle_{ph} ~ \tilde{\rho}_s(t) ~  _{ph}\langle m| H_I|n\rangle_{ph} 
\frac{e^{-\beta \omega_m}}{Z} e^{-i(\omega_n-\omega_m)\tau} 
 \nonumber \\
 ~~~~~~~~~~~~~~~~ + \left. \tilde{\rho}_s(t)~ _{ph}\langle n | H_I |m\rangle_{ph}~_{ph}\langle m| H_I|n\rangle_{ph} 
  \frac{e^{-\beta \omega_n}}{Z} e^{-i(\omega_n-\omega_m)\tau}
\right] .
\end{eqnarray}}
Next, we evaluate the first term in the above equation and show that it is zero at 
$T=0$. We observe that
{\begin{eqnarray}
 Tr_R[\tilde{H}_I(t)R_o]&=& \sum_{n}~ _{ph}\langle n |\tilde{H}_I(t)R_o|n \rangle_{ph} \nonumber \\
&=& 0.5 J  e^{-g^2}\sum_{l,j\neq l} \left[ e^{iH_s t} b^{\dagger}_l b_j e^{-iH_st}~ 
_{ph}\langle 0 |  \{\mathcal S^{{lj}^\dagger}_+ \mathcal S^{lj}_{-} 
 -1\}  |0 \rangle_{ph} \right] \nonumber \\ &=& 0 .
\end{eqnarray}}
Thus, we have $  \sum_{n}~ _{ph}\langle n| [\tilde{H}_I(t), \rho_s(0) \otimes R_o] |n \rangle_{ph}=0$
and the master equation at $T=0$ simplifies as:
\begin{eqnarray}
\fl~~~~~~~~~~~ \frac{d \tilde{\rho}_s(t)}{dt} 
&=& -\sum_{m} \int_0^\infty d\tau \left[ |_{ph}\langle 0 | H_I |m\rangle_{ph}|^2
 ~\tilde{\rho}_s(t)  e^{-i\omega_m\tau}  + \tilde{\rho}_s(t) ~ |_{ph}\langle 0 | H_I |m\rangle_{ph}|^2 
  e^{i\omega_m\tau} \right] \nonumber \\
 &&+ \sum_{n} \int_{0}^\infty d\tau \left[ _{ph}\langle n | H_I |0\rangle_{ph}  \tilde{\rho}_s(t) _{ph}\langle 0|
 H_I|n\rangle_{ph} e^{i\omega_n\tau} \right .
\nonumber \\
&& ~~~~~~~~~~~~~~~~~~ \left . +  _{ph}\langle n | H_I |0\rangle_{ph}  \tilde{\rho}_s(t) _{ph}\langle 0| H_I|n\rangle_{ph}
 e^{-i\omega_n\tau} \right]  \nonumber \\
&=& - \sum_n  
\left[ \int_0^{\infty} d\tau ~ e^{- i(\omega_n -i \eta)\tau} |_{ph}\langle 0 |H_I |n \rangle_{ph}|^2 ~ \tilde{\rho}_s(t)
\right . 
\nonumber \\
&& ~~~~~~~~~~ + \left .
\int_{0}^{\infty} d\tau ~ e^{i(\omega_n+i\eta)\tau}
 ~\tilde{\rho}_s(t) ~|_{ph}\langle 0 |H_I |n \rangle_{ph}|^2  \right .\nonumber \\
&&~~~~~~~~~~ - \left. \int_{-\infty}^{\infty} d\tau ~ e^{i\omega_n \tau} ~ _{ph}\langle n| H_I |0\rangle_{ph} 
~\tilde{\rho}_s(t)~ _{ph}\langle 0|H_I|n\rangle_{ph} 
    \right]. \nonumber \\
\label{21}
\end{eqnarray}
Now, we know $\int_{-\infty}^{\infty} d\tau e^{i \omega_n \tau} \propto \delta(\omega_n)$.
Therefore, on using this relation and  the fact that ${_{ph}\langle 0| H_I |0\rangle_{ph}} =0$, the 
third term in equation (\ref{21}) vanishes; hence, we get
\begin{eqnarray}
\frac{d \tilde{\rho}_s(t)}{dt} =  i~\sum_n \left[
  \frac{ |_{ph}\langle 0 |H_I |n \rangle_{ph}|^2 }{\omega_n} \tilde{\rho}_s(t) - 
\tilde{\rho}_s(t) \frac{ |_{ph}\langle 0 |H_I |n \rangle_{ph}|^2 }{\omega_n} 
\right]  .
\end{eqnarray}
The term $\sum_n \left[\frac{ |_{ph}\langle 0 |H_I |n \rangle_{ph}|^2 }{\omega_n} \right]$ 
corresponds to the effective Hamiltonian  $H^{(2)}$  in second order perturbation and commutes
 with $H_0$ (see section $3$). Let $|n\rangle_s$ be the simultaneous eigenstate for $H^{(2)}$ and $H_s$ with
 eigenvalues $E_n^{(2)}$ and $E_n^s$, respectively. Then, from the above equation we get:
\begin{eqnarray}
{_s\langle n|\tilde{\rho}_s(t) | m \rangle_s} &=&
 e^{-i(E^{(2)}_n- E_m^{(2)})t}~ _s\langle n|\tilde{\rho}_s(0) | m \rangle_s ,
\end{eqnarray}
which implies that
\begin{eqnarray}
 {_s\langle n|\rho_s(t) | m \rangle_s}
& =& e^{-i(E_n- E_m)t}~ _s\langle n|\rho_s(0) | m \rangle_s ,
\label{sol}
\end{eqnarray}
where $E_n= E_n^s+E_n^{(2)}$.
Thus we see from the above equation that there is only a phase shift but no decoherence!
{Since the matrix elements of an operator are invariant
under canonical transformation, it should be clear that no loss in off-diagonal density matrix elements 
(i.e., no decoherence)
 in the LF transformed 
frame of reference implies no loss in off-diagonal density matrix elements (i.e., no decoherence)
in the
original untransformed frame of reference.
Although the HCB's in the original frame of reference form polarons and are thus
entangled with the environment, nevertheless no decoherence results.
For greater clarity, the form of $_s\langle n|\rho_s(t) | m \rangle_s$ in the original frame of
reference and its associated non-decoherence is discussed in Appendix B
for a special two-spin case of IRHM. Thus, up to second order in perturbation, the 
 assumption $J^{\star} e^{-g^2} < < \omega$, the infinite range
of the Heisenberg model, and the Markov approximation 
 ($\tau_{c} \ll \tau_{s}$) together have ensured that the 
system, with a fixed  $S_T^{z}$, does not decohere.}

While the above analysis is valid in the regime $k_B T /\omega << 1$, the finite temperature
case  $k_B T /\omega \gtrsim 1$ needs additional extensive considerations
and will be dealt with elsewhere \cite{am_muz_sy2}.

\subsection{Decoherence due to Global Phonons:}
We will now analyze decoherence due to interaction of the spin system with global phonons.
To this end, we consider the following total Hamiltonian where all qubits of our IRHM interact identically
with the environment.
\begin{eqnarray}
\!\!\!\! H_{Tot} = H_{s}+ && \sum_i S^z_i \sum_{k} \omega_k (g_k a^{\dagger}_k + g_k^{\star}a_k) 
\nonumber \\
 && +  \sum_k \omega_k a^{\dagger}_k a_k ,
\label{Ham2}
\end{eqnarray}
where $H_s = H_{\rm IRHM} $ is the Hamiltonian of the system.
{(Here, for global phonons, since we do not use LF transformation,
we define $H_s$ as the untransformed system Hamiltonian.)} {Since the z-component of the 
total spin $S_{Total}^z$ (and thus the interaction Hamiltonian) commutes with $H_{\rm IRHM}$, the eigenstates
 having same eigenvalue $S_T^z$ constitute a DFS.
 To study the case when $S_T^z$ is not conserved and to obtain the form of the reduced density matrix $\rho_s(t)$, 
we study the dynamics of the system
through the following 
non-Markovian master equation}
\cite{YU}:
\begin{eqnarray}
\frac{d \rho_s(t)}{dt}
=&&
 -i[H_s,\rho_s(t)] + F(t)[L\rho_s(t),L] 
\nonumber \\
&&+F^{\star}(t)[L,\rho_s(t) L] , 
\end{eqnarray}
where 
 $L$ is the system operator
 that couples with the bath and satisfies the constraint $[L,H_s]=0$.
For the  total Hamiltonian in equation (\ref{Ham2}),
 $L = \sum_i S^z_i=S^z_{Total}$.
 Also, $ F(t)=\int_0^{t} \alpha(t-s)ds$ 
where $\alpha(t-s)= \eta(t-s)+i\nu(t-s)$ is the bath correlation function at temperature $T$ with
\begin{eqnarray}
\eta(t-s) &=&\sum_{k} |g_{k}|^2 \omega_k^2\coth(\frac{\omega_{k}}{2k_B T} )\cos[\omega_{k}(t-s)] , \nonumber \\
\nu(t-s) &=&-\sum_{k} |g_{k}|^2 \omega_k^2 \sin[\omega_{k}(t-s)] .
\end{eqnarray}
The function  $F(t)$ 
governs the non-Markovian dynamical features of the system.

Let $\{| n \rangle_s \}$ be the eigen basis  in which both the operators $S^z_{Total}$ 
and $H_s$ are simultaneously
diagonalized. Upon solving the master equation explicitly we get \cite{YU}:
\begin{eqnarray}
&&{_s\langle n|\rho_s (t)|m\rangle_s}
\nonumber \\   
 && ~~=  \exp \left (-i \left [(E_n^s -E_m^s)t + 
\left \{ (S^z_{Tn})^2-(S^z_{Tm})^2 \right \} Y(t) \right ] \right ) \nonumber \\
  && ~~~ \times \exp \left [- \left ( S^z_{Tn}-S^z_{Tm} \right )^2 X(t) \right ]
~{_s\langle n|\rho_s (0)|m\rangle_s} ,
\end{eqnarray}
where $E_n^s$ and $S^z_{Tn}$ are defined through
$H_{s}|n\rangle_s = E_n^s|n\rangle_s$ and $\sum_i S^z_i
|n\rangle_s =S^z_{Tn}|n\rangle_s$.
Furthermore, 
$X(t) \equiv \int_0^{t} F_R(s)ds $ and $Y(t) \equiv \int_0^{t} F_I(s)ds $ 
with $F_R(t) + i F_I(t) \equiv F(t)$.
{This implies  that,
when states $|m\rangle_s$ and $|n\rangle_s$ have the 
same z-component of the total spin $S^z_{T}$ ( i.e., $S^z_{Tn}=S^z_{Tm}$),
 the  matrix elements ${_s\langle n|\rho_s (t)|m\rangle_s}= 
{_s\langle n|\rho_s (0)|m\rangle_s} \exp [-i(E_n^s -E_m^s)t]$ 
display a decoherence free behaviour. But when $S^z_{T}$ is not conserved,
the off-diagonal matrix elements will diminish in general, i.e., the system undergoes decoherence.
In the language of HCBs, the eigenstates of the system 
with a fixed number of HCBs makeup a DFS. 
Furthermore, the entanglement entropy of the system will remain
unaltered since the density matrix evolves unitarily.
In future, using the above framework, we will consider the interesting case of
 dynamical evolution and decoherence
 of states with different 
 $S^z_T$ values.}

\section{ Discussion and Conclusions}
  In conclusion, we have shown that the eigenstates of $H_{eff}$ are the same as those of $H_{\rm IRHM}$
 and for Markov processes they
are decoherence free
under the coupling of the system to 
 local optical phonons. For global optical phonons (i.e., when 
all the qubits are exposed to the same  collective noise) the
eigenstates with the same $S_T^z$ form a DFS. 
A DFS is expected in the global phonon case because the Hamiltonian of the system
commutes with $\sum_i S^z_i$.
 But the important point is that, even in the local phonon case, it is still
 possible to  fully preserve coherence
 for the composite particle (i.e., polaronic HCB) system with a fixed number of particles. 
 {More specifically,  for local phonons, $_s\langle n|\rho_s(t) | m \rangle_s$ 
differs from $_s\langle n|\rho_s(0) | m \rangle_s$ only by a phase factor  
  and  $_s\langle n|\rho_s(0) | m \rangle_s$ can be obtained from 
{$_s\langle n|\rho_{\rm{IRHM}} | m \rangle_s$}
 (density matrix element of IRHM) by an exact unitary evolution \cite{gl}.}
 Later, we will analyze the non-Markov processes and see how the resultant dynamics deviates from  
the Markovian dynamics.

{ Earlier, a new type of resonating valence bond (RVB) states \cite{authors}
were constructed for four and six spins using
  homogenized linear superposition of the $S_T=0$ 
states of $H_{\rm IRHM}$; these RVB states have a  high bipartite entanglement.
The decoherence analysis in this paper is also applicable to these 
new RVB states which are groundstates of our $H_{\rm IRHM}$.
 Our RVB states are constructed using valence bond (VB) states which are
$S_T =0 $ states.
VB states are built from singlet states between pairs of spins. A general VB state 
is defined as:
\begin{eqnarray}
\!\!\!\!\!\!|\Psi\rangle_{\rm vb} =|\Phi_{i_1,j_1}\rangle \otimes |\Phi_{i_2,j_2}\rangle \otimes...\otimes 
|\Phi_{i_M,j_M} \rangle ,
\nonumber
\label{vb}
\end{eqnarray}
where  $ |\Phi_{i_k,j_k} \rangle \equiv 
\frac{1}{\sqrt2} \left (|\frac{1}{2}\rangle_{i_k} |-\frac{1}{2}\rangle_{j_k}
 -|-\frac{1}{2}\rangle_{i_k} |\frac{1}{2}\rangle_{j_k} \right )$ 
denotes the singlet dimer connecting a site $i_k$ in sub-lattice $A$ with a site $j_k$ in sub-lattice $B$.
Examples of our RVB states (that are constructed from spins 1, 2, 3, 4,... arranged
sequentially on the vertices of a regular polygon and that have 
 high bipartite entanglement) are $|\Psi_4\rangle_{\rm rvb}$ given below for four spins:
\begin{eqnarray}
\!\!\!\!\!\! |\Psi_4\rangle_{\rm rvb} \equiv \omega_3 (|\Phi_{1,2}\rangle
 \otimes |\Phi_{3,4}\rangle) 
+\omega_3^2(|\Phi_{1,4}\rangle
 \otimes |\Phi_{2,3}\rangle), 
 \label{psi_hs}
\end{eqnarray}
where $\omega_3$ ($=e^{i2\pi/3}$)  is a cube root of unity; and
$|\Psi_6\rangle_{\rm rvb}$ given below for six spins:
\begin{eqnarray}
 |\Psi_6\rangle_{\rm rvb} &=& \omega_4 (|\Phi_{1,2}\rangle \otimes |\Phi_{3,6} \rangle \otimes |\Phi_{4,5}\rangle)
 \nonumber \\
&+& \omega_4^2(| \Phi_{2,3} \rangle \otimes |\Phi_{1,4} \rangle \otimes |\Phi_{5,6}\rangle)
 \nonumber \\
&+&\omega_4^3(|\Phi_{1,6} \rangle \otimes |\Phi_{2,5}\rangle \otimes |\Phi_{3,4}\rangle) ,
\end{eqnarray}
where $\omega_4$  ($=e^{i2\pi/4}$) is a fourth root of unity.}

Before closing we will make a few general remarks.
Firstly, the $H_{\rm IRHM}$ model of this paper, deals with the
extreme case of distance independent interaction among the spins. On the other extreme end,
if one were to consider a nearest-neighbor interaction anisotropic Heisenberg chain [of the type
$ \sum_{i} \{J_{\perp} (S^x_i S^x_{i+1} + S^y_i S^y_{i+1}) + J_{\parallel} S^z_i S^z_{i+1}\}$ 
where $J_{\perp} > J_{\parallel} > 0$]
with a strong coupling to local phonons [introducing the additional terms
 $g \omega \sum_i S^z_i (a^{\dagger}_i + a_i) 
+  \omega \sum_i  a^{\dagger}_i a_i$], { then the spin system undergoes a Luttinger liquid to a 
spin-density-wave transition upon turning on the spin-phonon interaction 
and  decoheres} \cite{am_muz_sy}. In general, distance-dependent-interaction in spin Hamiltonians will
fall somewhere in between the above two extreme cases.

Next, our decoherence analysis for local optical phonons will continue to be valid
 even for the more general optical phonon terms given below:
\begin{eqnarray}
\frac{1}{N^{1/2}}\sum_{i,k} S^z_i 
 [ \omega_k (g_k a^{\dagger}_{k,i} + g_k^{\star}a_{k,i})] 
 +  \sum_{k,i} \omega_k a^{\dagger}_{k,i} a_{k,i} .
\end{eqnarray}
We also must mention that our approach cannot accommodate the acoustic phonon case as here 
the condition $J^{\star}e^{-g^2} << \omega_k$
 cannot be satisfied in the long wavelength limit.

\ack
{ One of the authors (S. Y.) would like to thank 
G. Baskaran, R. Simon, S. Ghosh, and S. Reja for valuable discussions.}

\appendix
\section{}

\begin{figure}[]
\begin{center}
\includegraphics[width=3.5in,height=3.0in]{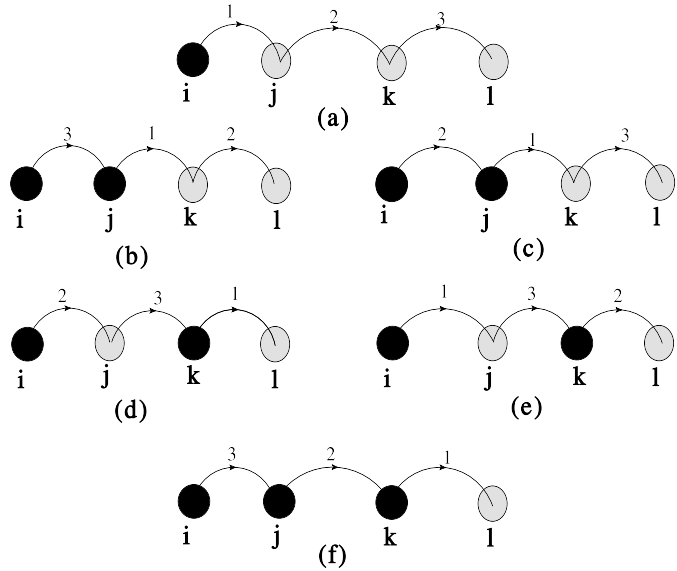}
\caption{ Open loop hopping processes contributing to effective hopping term $T_n^{li}$ in
third-order 
perturbation theory.
Here empty circles
correspond to sites with no particles while filled circles correspond 
to sites with hard-core-bosons. The numbers 1, 2, and 3 indicate the order of
hopping.
}
\label{fey1}
\end{center}
\end{figure}
\begin{figure}[]
\begin{center}
\includegraphics[width=1.5in,height=3.0in]{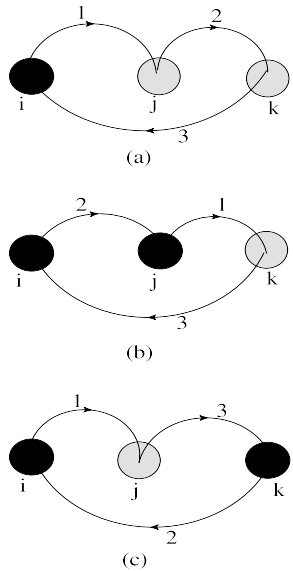}
\caption{Closed-loop hopping processes contributing to effective interaction term $V_n^i$ in
third-order 
perturbation theory.
Here filled (empty) circles 
correspond to sites with (without) hard-core-bosons. The numbers 1, 2, and 3 represent hopping sequence.}
\label{fey2}
\end{center}
\end{figure}
\begin{figure}[]
\begin{center}
\includegraphics[width=1.3in,height=3.0in]{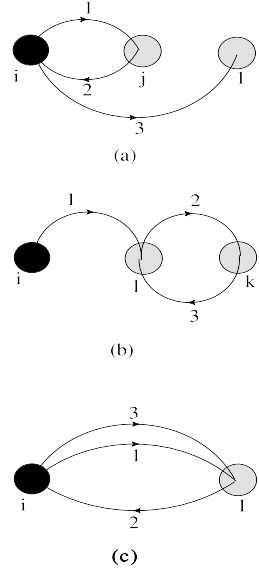}
\caption{Hopping processes (involving closed loops) contributing to effective hopping term $T_{Cn}^{li}$ in
third-order 
perturbation theory. Filled (empty) circles represent occupied (unoccupied) sites.
}
\label{fey3}
\end{center}
\end{figure}

In this appendix, we will show that the third-order
perturbation theory also produces a term that has the same
eigenstates as IRHM. To this end, we obtain the following third-order
 perturbation term in the effective Hamiltonian:
\begin{eqnarray}
\!\!\!H^{(3)}\!=\sum_{m\neq 0,n \neq 0}\!\!\!\! \frac{{_{ph}}\!\langle0|H_{I}|m\rangle_{ph}
~{_{ph}}\!\langle m|H_{I}|n\rangle_{ph}
~{_{ph}}\!\langle n|H_{I}|0\rangle_{ph}}
{{\Delta E_m^{ph}}{\Delta E_n^{ph}}}  . 
\nonumber \\ 
\label{H3}
\end{eqnarray}
Here $\Delta E_{m}^{ph}= \omega_{m}-\omega_{0}$.
Evaluation of $H^{(3)}$ leads to various hopping terms and interaction terms.
\begin{eqnarray}
H^{(3)}=\sum_{i,l\neq i} \left [\sum_{n=1}^{6} t_n T_n^{li} 
   + \sum_{n=1}^3 t_{cn} T_{Cn}^{li} \right ] + \sum_i \sum_{n=1}^3 v_n V^i_n , 
\nonumber \\ 
\label{tvtc}
\end{eqnarray}
where $t_n \sim (J^3 e^{-g^2})/(g^2 \omega)^2$,
 $t_{cn} \sim J^3 e^{-g^2}/(g\omega)^2$, and  $v_n \sim J^3 /(g^2 \omega)^2$
(as will be explained later).
We will demonstrate below that  $H^{(3)}$ is of the following form
\begin{eqnarray}
 H^{(3)} =\sum_{i,l > i} \left [ T(\sum_k n_k) b^{\dagger}_l b_i + {\rm H.c.} \right ] +  \sum_{i} V(\sum_k n_k) n_i  , 
\nonumber \\ 
\label{H3_form}
\end{eqnarray}
where $T$ and $V$ are functions of the total number operator $\sum_k n_k$.
 Since the IRHM commutes with the total number operator, $H^{(3)}$ has the same eigenstates as IRHM! 

 There are six open-loop hopping processes $T_n^{li}$ depicted in {figure} \ref{fey1}.  We analyze
them sequentially below.
\begin{eqnarray}
\!\!\!\!\!\!\!\!\!\!\!\!\!\!\!\!T_1^{li}
 &=& \sum_{k \neq i,l,j}\sum_{j \neq i,l} b^{\dagger}_l b_k b^{\dagger}_k b_j b^{\dagger}_j b_i 
\nonumber \\
 &=& \sum_{k \neq i,l,j} (1-b^{\dagger}_k b_k) \sum_{j \neq i, l} (1-b^{\dagger}_j b_j) b^{\dagger}_l b_i 
\nonumber \\
&=& \left [\sum_{k \neq i,l} (1-b^{\dagger}_k b_k) -1 \right ] \left [ \sum_{j \neq i, l} 
(1-b^{\dagger}_j b_j)\right ] b^{\dagger}_l b_i 
\nonumber \\
&=&\left [ \sum_{k \neq i,l} (1-b^{\dagger}_k b_k)-1 \right ] \left [ (N-2) - 
\sum_{j \neq l} b^{\dagger}_j b_j \right ] b^{\dagger}_l b_i
\nonumber \\
&=& \left [\sum_{k \neq i,l} (1-b^{\dagger}_k b_k)-1 \right ] \left [(N-1) - \sum_{j } b^{\dagger}_j b_j \right ] b^{\dagger}_l b_i
\nonumber \\
&=& \left [(N-1) - \sum_{j } b^{\dagger}_j b_j \right ] 
\left [\sum_{k \neq i,l} (1-b^{\dagger}_k b_k)-1 \right ]b^{\dagger}_l b_i
\nonumber \\
&=&\left [(N-1) - \sum_{j } b^{\dagger}_j b_j \right ] 
\left [(N-2) -\sum_{k } b^{\dagger}_k b_k \right ]b^{\dagger}_l b_i .
\nonumber \\
\end{eqnarray}
The second hopping
process $T_2^{li}$ in {figure} \ref{fey1} (b)  is given by 
\begin{eqnarray}
\!\!\!\!\!\!\!\!\!\!\!\!\!\!\!\!T_2^{li}
 &=& \sum_{k \neq i,l,j}\sum_{j \neq i,l} b^{\dagger}_j b_i b^{\dagger}_l b_k b^{\dagger}_k b_j 
\nonumber \\
 &=& \sum_{k \neq i,l,j} (1-b^{\dagger}_k b_k) \sum_{j \neq i, l} b^{\dagger}_j b_j b^{\dagger}_l b_i 
\nonumber \\
&=& \sum_{k \neq i,l} (1-b^{\dagger}_k b_k) \sum_{j \neq i, l} 
b^{\dagger}_j b_j b^{\dagger}_l b_i 
\nonumber \\
&=&\sum_{k \neq i,l} (1-b^{\dagger}_k b_k) \left [ 
\sum_{j } b^{\dagger}_j b_j -1 \right ] b^{\dagger}_l b_i
\nonumber \\
&=& \left [ 
\sum_{j } b^{\dagger}_j b_j -1 \right ]\left [ (N-1)-\sum_{k}b^{\dagger}_k b_k) \right ] b^{\dagger}_l b_i .
\end{eqnarray}
The hopping process $T_3^{li}$ in {figure} \ref{fey1} (c)  
is expressed as $T_3 ^{li}=\sum_{k \neq i,l,j}\sum_{j \neq i,l} b^{\dagger}_l b_k b^{\dagger}_j b_ib^{\dagger}_k b_j
 =T_2^{li}$.
The fourth hopping process $T_4^{li}$  in {figure} \ref{fey1} (d) is obtained as follows.
\begin{eqnarray}
\!\!\!\!\!\!\!\!\!\!\!\!\!\!\!\!T_4^{li}
 &=& \sum_{j \neq i,l, k}\sum_{k \neq i,l} b^{\dagger}_k b_j b^{\dagger}_j b_i b^{\dagger}_l b_k 
\nonumber \\
 &=& \sum_{j \neq i,l,k} (1-b^{\dagger}_j b_j) \sum_{k \neq i, l} b^{\dagger}_k b_k b^{\dagger}_l b_i 
\nonumber \\
&=& T_2^{li}  .
\end{eqnarray}
The hopping process $T_5^{li}$  in {figure} \ref{fey1} (e) yields
 $T_5^{li} =\sum_{j \neq i,l, k}\sum_{k \neq i,l} b^{\dagger}_k b_j  b^{\dagger}_l b_k b^{\dagger}_j b_i 
=T_4^{li}$.
 We  analyze below the last hopping 
process $T_6^{li}$ in {figure} \ref{fey1} (f).
\begin{eqnarray}
\!\!\!\!\!\!\!\!\!\!\!\!\!\!\!\!T_6^{li}
 &=& \sum_{k \neq i,l,j}\sum_{j \neq i,l} b^{\dagger}_j b_i b^{\dagger}_k b_j b^{\dagger}_l b_k 
\nonumber \\
 &=& \sum_{k \neq i,l,j} b^{\dagger}_k b_k \sum_{j \neq i, l} b^{\dagger}_j b_j b^{\dagger}_l b_i 
\nonumber \\
&=& \left [\sum_{k \neq i,l} b^{\dagger}_k b_k - 1 \right ] \sum_{j \neq i, l} 
b^{\dagger}_j b_j b^{\dagger}_l b_i 
\nonumber \\
&=&\left [\sum_{k \neq i,l} b^{\dagger}_k b_k - 1 \right ]\left [ 
\sum_{j } b^{\dagger}_j b_j -1 \right ] b^{\dagger}_l b_i
\nonumber \\
&=& \left [ 
\sum_{j } b^{\dagger}_j b_j -1 \right ]\left [ \sum_{k}b^{\dagger}_k b_k-2 \right ] b^{\dagger}_l b_i  .
\end{eqnarray}

We will now deal with closed-loop hopping processes such as those in {figure} \ref{fey2}.
These lead to effective interactions. The process $V_1^i$ in {figure} \ref{fey2} (a),  obtained from
figure \ref{fey1} (a) by setting $l=i$, is given as follows. 
\begin{eqnarray}
\!\!\!\!\!\!\!\!\!\!\!\!\!\!\!\!V_1^i
 &=& \sum_{k \neq i,j}\sum_{j \neq i} b^{\dagger}_i b_k b^{\dagger}_k b_j b^{\dagger}_j b_i 
\nonumber \\
 &=& \sum_{k \neq i,j} (1-b^{\dagger}_k b_k) \sum_{j \neq i} (1-b^{\dagger}_j b_j) b^{\dagger}_i b_i 
\nonumber \\
&=& \left [\sum_{k \neq i} (1-b^{\dagger}_k b_k) -1 \right ] \left [ \sum_{j \neq i} 
(1-b^{\dagger}_j b_j)\right ] b^{\dagger}_i b_i 
\nonumber \\
&=&\left [(N) - \sum_{j } b^{\dagger}_j b_j \right ] 
\left [(N-1) -\sum_{k } b^{\dagger}_k b_k \right ]b^{\dagger}_i b_i .
\end{eqnarray}
Next, the hopping process $V_2^i$ corresponding to closed loop in {figure} \ref{fey2} (b)
 is obtained from {figure} \ref{fey1} (c) by taking $l=i$.
\begin{eqnarray}
\!\!\!\!\!\!\!\!\!\!\!\!\!\!\!\!V_2^i
 &=& \sum_{k \neq i,j}\sum_{j \neq i} b^{\dagger}_i b_k b^{\dagger}_j b_i b^{\dagger}_k b_j 
\nonumber \\
 &=& \sum_{k \neq i,j} (1-b^{\dagger}_k b_k) \sum_{j \neq i} b^{\dagger}_j b_j b^{\dagger}_i b_i 
\nonumber \\
&=& \sum_{k \neq i} (1-b^{\dagger}_k b_k) \sum_{j \neq i} 
b^{\dagger}_j b_j b^{\dagger}_i b_i 
\nonumber \\
&=&\sum_{k \neq i} (1-b^{\dagger}_k b_k) \left [ 
\sum_{j } b^{\dagger}_j b_j -1 \right ] b^{\dagger}_i b_i
\nonumber \\
&=& \left [ 
\sum_{j } b^{\dagger}_j b_j -1 \right ]\left [ (N)-\sum_{k}b^{\dagger}_k b_k) \right ] b^{\dagger}_i b_i .
\end{eqnarray}
Lastly, the hopping $V_3^i$ [depicted by the closed loop in {figure} \ref{fey2} (c)]
 is obtained from {figure} \ref{fey1} (e) by setting $l=i$.
\begin{eqnarray}
\!\!\!\!\!\!\!\!\!\!\!\!\!\!\!\!V_3^i
 &=& \sum_{j \neq i, k}\sum_{k \neq i} b^{\dagger}_k b_j b^{\dagger}_i b_k b^{\dagger}_j b_i
\nonumber \\
 &=& \sum_{j \neq i,k} (1-b^{\dagger}_j b_j) \sum_{k \neq i} b^{\dagger}_k b_k b^{\dagger}_i b_i 
\nonumber \\
&=& V_2^i  .
\end{eqnarray}

Finally, we consider {figures} \ref{fey3} (a), (b), and (c) 
 which deal with effective hopping terms $T_{Cn}^{li}$ involving closed loops. The effective hopping
 term $T_{C1}^{li}$,
corresponding to {figure} \ref{fey3} (a), is obtained by setting $k=i$ in {figure} \ref{fey1} (a):
\begin{eqnarray}
\!\!\!\!\!\!\!\!\!\!\!\!\!\!\!\!T_{C1}^{li}
 &=& \sum_{j \neq i, l} b^{\dagger}_l b_i b^{\dagger}_i b_j b^{\dagger}_j b_i 
\nonumber \\
 &=&  \sum_{j \neq i, l} (1-b^{\dagger}_j b_j) b^{\dagger}_l b_i 
\nonumber \\
&=& \left [ (N-2) - 
\sum_{j \neq l} b^{\dagger}_j b_j \right ] b^{\dagger}_l b_i
\nonumber \\
&=&  \left [(N-1) - \sum_{j } b^{\dagger}_j b_j \right ] b^{\dagger}_l b_i .
\end{eqnarray}
To obtain the effective hopping term $T_{C2}^{li}$
corresponding to {figure} \ref{fey3} (b), we take $j=l$ in {figure} \ref{fey1} (a):
\begin{eqnarray}
\!\!\!\!\!\!\!\!\!\!\!\!\!\!\!\!T_{C2}^{li}
 &=& \sum_{k \neq i,l}
b^{\dagger}_l b_k b^{\dagger}_k b_l b^{\dagger}_l b_i 
\nonumber \\
 &=& \sum_{k \neq i,l} (1-b^{\dagger}_k b_k)  b^{\dagger}_l b_i 
\nonumber \\
&=& \left [ (N-2) - 
\sum_{k \neq l} b^{\dagger}_k b_k \right ] b^{\dagger}_l b_i
\nonumber \\
&=& \left [(N-1) - \sum_{k } b^{\dagger}_k b_k \right ] b^{\dagger}_l b_i 
\nonumber \\
&=& T_{C1}^{li} .
\end{eqnarray}
The effective hopping term $T_{C3}^{li}$ depicted in {figure} \ref{fey3} (c) 
[upon setting $k=i$ and $j=l$ in {figure} \ref{fey1} (a)] is given by
\begin{eqnarray}
\!\!\!\!\!\!\!\!\!\!\!\!\!\!\!\!T_{C3}^{li}
 = b^{\dagger}_l b_i b^{\dagger}_i b_l b^{\dagger}_l b_i 
= b^{\dagger}_l b_i .
\end{eqnarray}

 Thus we have shown that $H^{(3)} $ contains effective hopping terms 
{$\sum_{i,l>i} [T(\sum_k n_k) b^{\dagger}_l b_i$ \\ $+ {\rm H.c.]} $}
 and effective interaction terms ($\sum_{i} V(\sum_k n_k) n_i$). Since 
 $T$ and $V$ are functions of the total number operator,
 $H^{(3)}$ and IRHM have the same eigenstates. These arguments can be extended to even
higher-order perturbation theory to show that the effective Hamiltonian (after taking all
orders of perturbation into account) will give the same eigenstates as IRHM!

\begin{figure}[]
\begin{center}
\includegraphics[width=1.5in,height=3.0in]{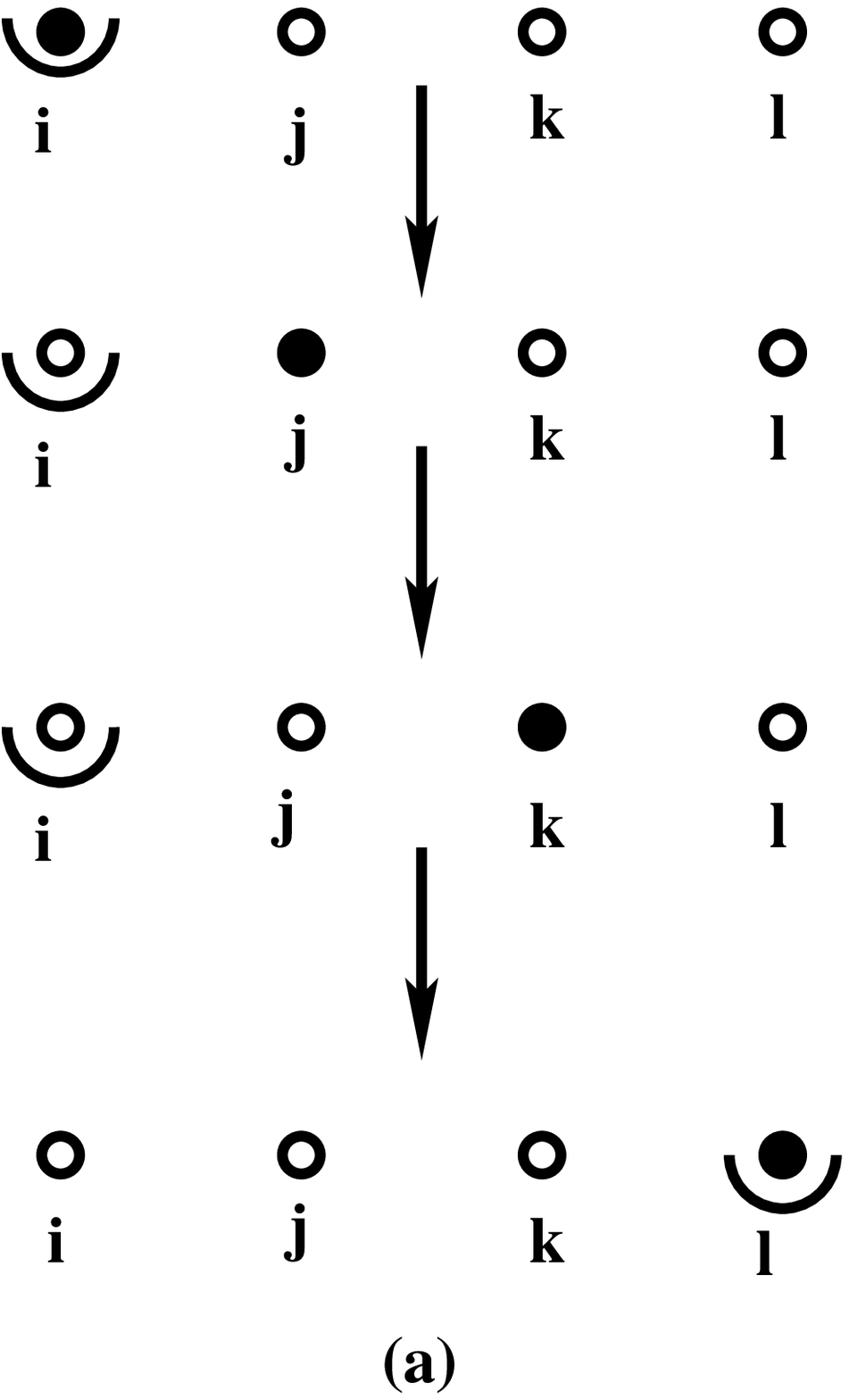}\\
\vspace{.3cm}
\includegraphics[width=2.5in,height=2.0in]{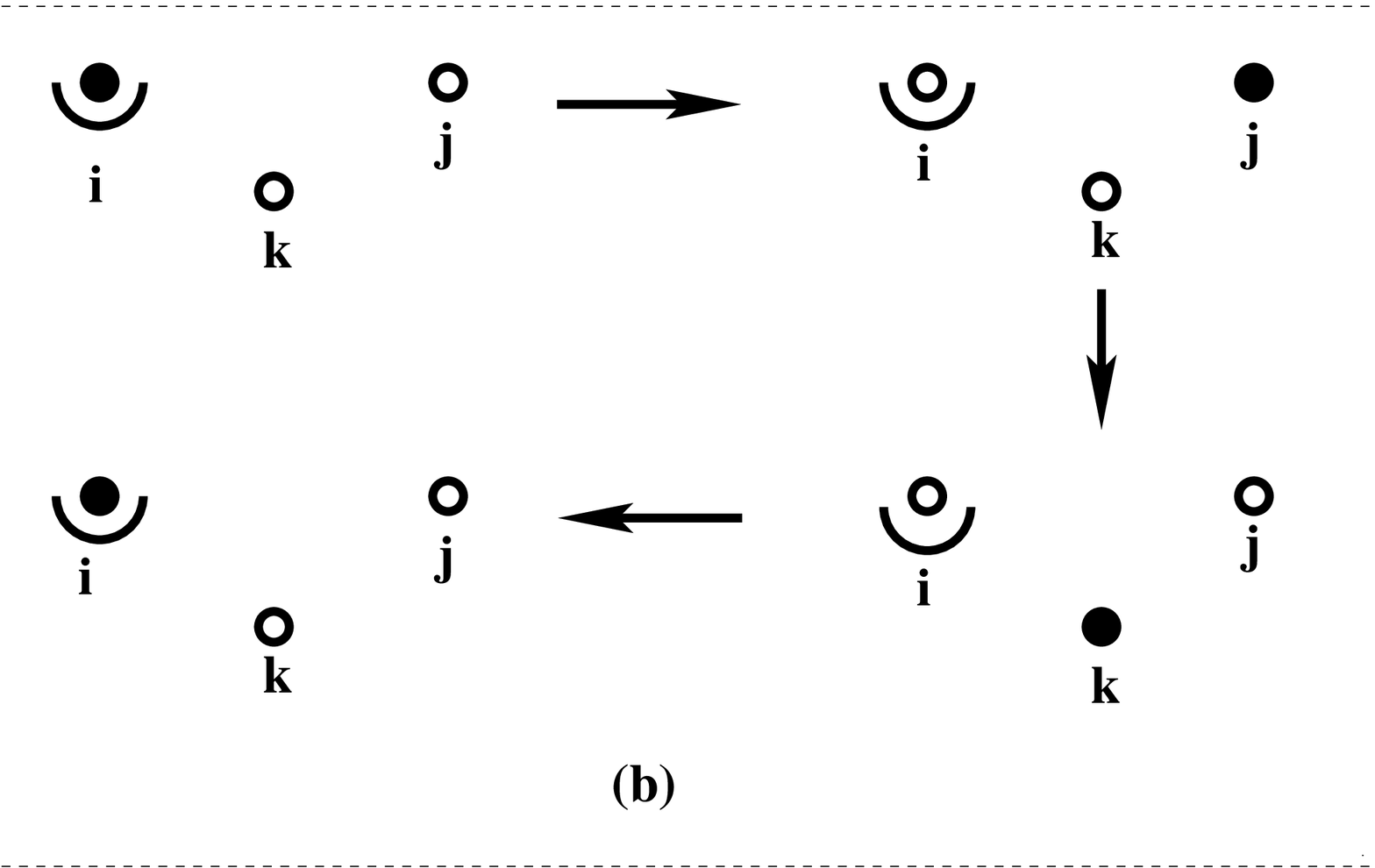}\\
\vspace{.3cm}
\includegraphics[width=2.5in,height=2.0in]{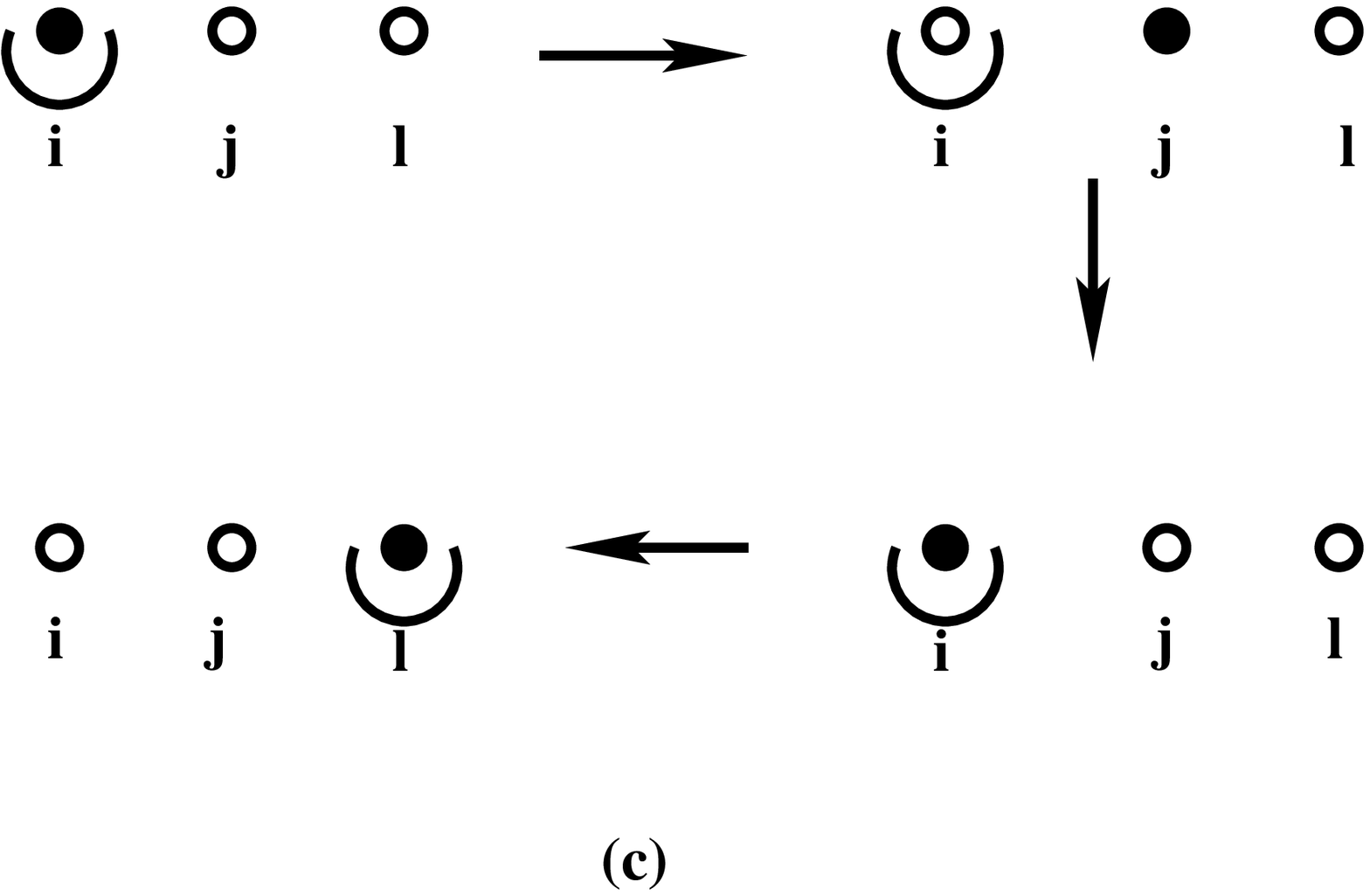}
\caption{Schematic diagrams (a), (b), and (c), corresponding
to the hopping processes depicted in {figure} \ref{fey1} (a), {figure} \ref{fey2} (a), 
and {figure} \ref{fey3} (a), respectively,  yield coefficients
$t_n$, $v_n$, and $t_{cn}$, respectively.  The intermediate states give the typical dominant
contributions. Here empty circles correspond to empty sites, while filled circles
 indicate particle positions. Parabolic curve
at a site depicts full distortion at that site with
corresponding energy $-g^2 \omega$ ($+g^2 \omega$) if the hard-core-boson is present
(absent) at that site.
}
\label{schemat}
\end{center}
\end{figure}
We will now explain the expressions for the coefficients $t_n$, $v_n$, and $t_{cn}$
in equation (\ref{tvtc}), 
 obtained from third-order
perturbation theory, using typical schematic
diagrams shown in {figure} \ref{schemat} [for details of corresponding
diagrams and analysis in second order perturbation, see reference \cite{srsypbl}]. We consider two distinct time scales
associated with  hopping processes between two sites:
(i) $\sim 1/(Je^{-g^2})$ corresponding to either full distortion at a site
to form a small polaronic potential well (of energy $-g^2 \omega$) or
full relaxation from the small polaronic distortion and (ii) $\sim 1/J$
related to negligible distortion/relaxation at a site. 
The coefficient $t_n$ corresponds
to the typical dominant distortion processes shown schematically in {figure} \ref{schemat} (a)
with the pertinent typical  hopping processes being depicted in
{figure} \ref{fey1} (a). In {figure} \ref{schemat} (a), after the HCB hops
 away from the initial site, the intermediate states
have the same distortion as the initial state. Next,
when the HCB hops to its final site there is a distortion
at this final site with a concomitant relaxation at the initial site.
Hence the contribution to the coefficient $t_n$ becomes
$J/(2 g^2 \omega) \times J/(2 g^2 \omega) \times J e^{-g^2} \sim J^3 e^{-g^2}/(g^2 \omega)^2$.
As regards coefficient $v_n$, it can be deduced based on the typical dominant 
hopping-cum-distortion 
processes depicted in {figure} \ref{schemat} (b) which typifies the hopping processes in {figure} \ref{fey2} (a).
In {figure} \ref{schemat} (b), when the particle hops to different sites and reaches finally the initial site,
there is no change in distortion at any site. Hence $v_n$ can be estimated to be
$J/(2 g^2 \omega) \times J/(2 g^2 \omega) \times J  \sim J^3 /(g^2 \omega)^2$.
Lastly, we obtain the coefficient $t_{cn}$ by considering the 
typical dominant diagram in {figure} \ref{schemat} (c)
corresponding to the typical process in {figure} \ref{fey3} (a). In {figure} \ref{schemat} (c),
where the first intermediate state depicts the particle hopping but leaving the distortion unchanged,
we get a contribution $J/(2g^2 \omega)$; for the next intermediate state, where the HCB returns to
the initial site, the initial site has to undergo a slight relaxation (involving absorbing a phonon
so as to yield a non-zero denominator in the perturbation theory) leading to the contribution $J/\omega$;
and lastly, when the HCB hops to the final site, there is a distortion
at the final site with a simultaneous relaxation at the initial site thereby producing
a contribution $J e^{-g^2}$. Thus we calculate $t_{cn}$ to be
 $J/(2g^2 \omega) \times J/\omega \times J e^{-g^2} \sim J^3 e^{-g^2}/(g \omega)^2$ \cite{sy1}.
 
\section{}

In equation (\ref{sol}), the matrix element $_s\langle n|\rho_s(t) | m \rangle_s$ can be written as
{\begin{eqnarray}
 _s\langle n|\rho_s(t) | m \rangle_s  
&=& _s\langle n|~\Bigg[\sum_{n}~ _{ph}\langle n| \rho_T(t) | n \rangle_{ph}\Bigg]~| m \rangle_s \nonumber \\
&=&~  _s\langle n|\sum_{n}~_{ph}\langle n| e^S \rho^o_T(t)
    e^{-S} | n \rangle_{ph}~| m \rangle_s ,
 \nonumber \\
\label{den}
\end{eqnarray}}
where $\rho^{o}_T(t)$ is the total density matrix in the original frame of reference .
Now, we illustrate this quantity by considering the simple two-spin (i.e., N=2) case of the IRHM.
The singlet state $\frac{1}{\sqrt{2}}(| \uparrow \downarrow \rangle - | \downarrow \uparrow \rangle)$
 and the triplet state  
$\frac{1}{\sqrt{2}}(| \uparrow \downarrow \rangle + | \downarrow \uparrow \rangle)$
are the  $S^z_T=0$ eigenstates
of the two-qubit IRHM Hamiltonian; in HCB language, these states are expressed as
$\frac{1}{\sqrt{2}}(| 10 \rangle - | 01 \rangle)$ 
and $\frac{1}{\sqrt{2}}(| 10 \rangle + | 01 \rangle)$, respectively.
Now, the operator $e^{-S}$ can be expressed as
{\begin{eqnarray}
\fl e^{-S} &=& e^{g\sum_{i=1,2}(n_i-\frac{1}{2}) (a_i - a^\dagger _i)} 
= \prod_{i=1,2} e^{g(n_i-\frac{1}{2})(a_i - a^\dagger _i)}
= \prod_{i=1,2} \Bigg[n_iX_i + (1-n_i)X_i^{\dagger}\Bigg] ,
\end{eqnarray}}
where $X_i=e^{\frac{g}{2}(a_i - a^\dagger _i)}$.
Using the above, we obtain
\begin{eqnarray}
 e^{-S}\frac{1}{\sqrt{2}}(| 10 \rangle \pm | 01 \rangle) | m_1, m_2 \rangle_{ph}
 = [{X}_1 {X}^{\dagger}_2 |10\rangle \pm {X}_2 {X}^{\dagger}_1  |01\rangle ]| m_1, m_2 \rangle_{ph} . \nonumber \\
\end{eqnarray}
{where $m_1$ and $m_2$ correspond to phonon occupation numbers at site $1$ and site $2$ respectively.}
Therefore, from equation (\ref{den}) we can write the density matrix element between singlet and 
triplet states in the original frame of reference as
{\begin{eqnarray}
\fl \frac{1}{2} \left ( \langle10 | -  \langle01 | \right ) \rho_{s}(t) 
\left ( | 10 \rangle + | 01 \rangle \right ) \\
\fl ~ =\frac{1}{2}\sum_{m_1,m_2} {_{ph}}\langle m_1,m_2 | 
\left ( \langle10 | {X}_{2}{X}^{\dagger}_1 - 
 \langle01 | {X}_{1}{X}^{\dagger}_2 \right )
  \rho^o_{T}(t) 
  \left ( {X}_1 {X}^{\dagger}_2| 10 \rangle
 + {X}_2 {X}^{\dagger}_1| 01 \rangle \right ) | m_1,m_2 \rangle_{ph} .
 \nonumber \\
\label{off-dia}
\end{eqnarray}}
Depending upon the presence or absence of HCB, appropriate
deformation will be produced at each site and
$\left [\left ({X}_1{X}^{\dagger}_2| 10 \rangle \pm
{X}_2 {X}^{\dagger}_1| 01 \rangle \right )| m_1,m_2 \rangle_{ph} \right ]$ represents polaronic states.
Furthermore, in equation (\ref{off-dia}),
no loss in the off-diagonal matrix element on the left hand side 
implies no loss in the off-diagonal matrix element on the right hand side (i.e., 
no loss in the
measured density matrix elements in the original frame of reference) which in turn means no decoherence results.
\\
\\
\\
{$^*$ {\rm Contributed equally to this work.}}


\begin{thebibliography}{47}
\bibitem{indexk}
The bath at each site is $\omega \sum_{k,j} a^{\dagger}_{k,j} a_{k,j}$ and $a_{k,j}$
is the phonon destruction operator. Here,
we suppressed the index $k$ (which represents the phonon modes) as it does not change
any results.
\bibitem{gl} If the perturbation $(H_s-H_{\rm IRHM})$ is adiabatically turned on ( i.e.,
 $H_s(t)=H_{\rm IRHM}+(H_s-H_{\rm IRHM})e^{-\lambda| t |}$, 
where $\lambda$ is a small positive quantity), the eigenstate $| \psi\rangle$ of our starting Hamiltonian $H_s$ can be unitarily connected to 
the eigenstate $| \phi\rangle$ of $H_{\rm IRHM}$,
 i.e., $| \psi(0)\rangle=U_I^{\dag}(-\infty)| \phi\rangle$, 
where $U_I(t)=e^{-i\int\limits_0^t(H_s-H_{\rm IRHM})e^{-\lambda | t' |}dt'}$.
\bibitem{sm_par}
For a large N system, the largest energy contribution due to terms involving closed loops is intensive. The dominant
contribution, in second and higher order perturbation,
comes from open loops such as those in {figure} 1. The approximate open loop terms in zeroth, second,
and third order perturbation are, respectively, $Je^{-g^2} \sum_{i,j\neq i} b^{\dagger}_i b_j$, $[NJ^2 e^{-g^2}/(g^2 \omega)]
\sum_{i,j\neq i} b^{\dagger}_i b_j$,
and $[N^2 J^3 e^{-g^2}/(g^2 \omega)^2 ]\sum_{i,j \neq i} b^{\dagger}_i b_j$; then, the small parameter
is $J^{\star}/(g^2 \omega)$.
\end{thebibliography}
\end{document}